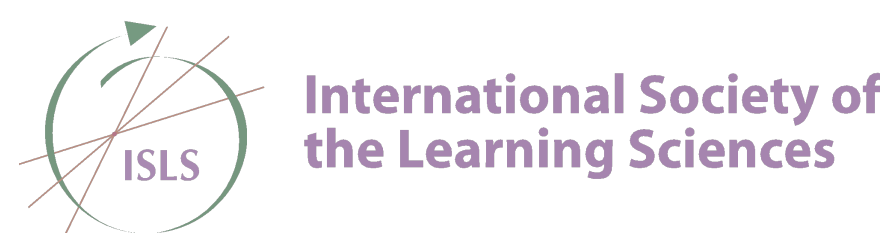

# An Activity-Theoretical Approach to Teacher Professional Development in Pedagogical AI Agent Design

Haiyang Xin, Qiannan Niu, Shuang Li, Yimeng Sun
Tony@cocorobo.cc, niuqiannan@cocorobo.cc, lishuang@cocorobo.cc, sunyimeng@cocorobo.cc
COCOROBO Limited
Ching Sing Chai, The Chinese University of Hong Kong, CSChai@cuhk.edu.hk
Lingyun Huang, The Education University of Hong Kong, lingyunhuang@eduhk.hk
Gaowei Chen, The University of Hong Kong, gwchen@hku.hk

**Abstract:** This two-cycle formative intervention study examined why teachers disengage from AI agent creation after professional development—a low engagement paradox—and tested whether systemic redesign could address it. Cycle 1 (*N*=218) revealed that despite completing comprehensive TPD, 87% of teachers ceased creating within three weeks, with behavioral tracking and interview analysis identifying systemic contradictions as the source of psychological need frustration rather than capacity deficits. Cycle 2 (*N*=26) implemented Cultural-Historical Activity Theory and Self-Determination Theory-driven redesign directly targeting diagnosed contradictions, achieving synchronized enhancement of both capacity and willingness. The findings reframe implementation failure as a rational response to need-thwarting systems and offer a replicable CHAT-SDT diagnostic framework for transformative professional development.

## Introduction

The integration of generative artificial intelligence (GenAI) into education hold substantial promise for creating personalized learning experiences and adaptive learning pathways (Pesovski et al., 2024), which may reduce teacher workload while enhancing instructional efficiency through automating routine tasks such as lesson planning and assessment grading (Chaudhry & Kazim, 2022). However, direct introduction of general-purpose AI tools has revealed critical challenges from both instructional and technical perspectives. From an instructional perspective, overreliance on AI may erode teachers' professional judgment by displacing skilled epistemic actions and subordinating epistemic agency to efficiency (Chen, 2025). Furthermore, inherent limitations of large language models—including hallucinations, embedded biases, and context-blending vulnerabilities—necessitate sustained human oversight (Huang et al., 2025).

To meaningfully harness generative AI, teachers must evolve from passive tool users to active designers of pedagogical AI agents (PA) according to their instructional goals and student learning needs (Lan & Chen, 2024). This transformation requires integrated understanding of how technology, pedagogy, and subject matter intersect in AI-enhanced environments—technological pedagogical content knowledge (TPACK; Koehler & Mishra, 2009)—alongside ethical reasoning and pedagogical design agency. However, two critical gaps hinder this transformation**.** First, existing pedagogical agent research has predominantly focused on student-facing applications and their learning effects—such as motivation, affective support, and comprehension (Ortega-Ochoa et al., 2024; Sikström et al., 2024). Little exploration has looked into how teachers can actively refine and adapt these agents to align with their instructional goals and contextual needs (Chen et al., 2025). Secondly, professional development research has been criticized for employing simplistic conceptualizations that isolate skill training from the complex intersections of teachers' knowledge, beliefs, and institutional cultures (Ertmer & Ottenbreit-Leftwich, 2010; Opfer & Pedder, 2011). This limitation intensifies in AI pedagogical agent creation, where teachers consistently report insufficient training, lack of institutional support, and knowledge gaps in understanding AI's educational affordances (Lan & Chen, 2024). Yet existing PD models adopt a process-product logic that views teacher learning as following more or less directly from the frequency of specific training activities (Opfer & Pedder, 2011), implicitly assuming that enhanced technological capacity automatically generates sustained design practice, an assumption that overlooks the motivational and systemic factors determining whether teachers will actively integrate AI agents beyond initial workshops.

Educational systems constitute complex adaptive systems where conventional single-dimension training models prove inadequate for addressing rapid technological iteration and structural constraints. To investigate this challenge specifically in the context of AI pedagogical agent creation, we conducted two-cycle intervention research employing Cultural-Historical Activity Theory and Self-Determination Theory. We began with an initial hypothesis that structuring teacher professional development (TPD) to enhance teachers' TPACK with continuous support, would sustain PA creation practice. We then investigated the sustainability of this effect and responded to emergent patterns through the Cycle 2 redesign. Our inquiry was organized around three interconnected questions:

- RQ1: How do teachers' behavioral engagement patterns in AI agent creation develop following baseline professional development?
- RQ2: What systemic and psychological mechanisms shape teachers' willingness to sustain AI agent creation practice? How do activity system contradictions influence teachers' motivational dynamics?
- RQ3: Can SDT-driven activity system redesign, directly targeting diagnosed contradictions, simultaneously enhance teachers' capacity and willingness for sustained AI agent creation?

## Theoretical Framework

### Cultural-Historical Activity Theory (CHAT)

This study adopts Engeström's (2001) activity theory to analyze teachers' AI agent creation as an activity system comprising six interrelated elements: subject, object, tools, rules, community, and division of labor. Contradictions serve as the internal driving force for systemic development, representing historically accumulating structural tensions rather than superficial conflicts. We focus on three contradiction types: primary contradictions (Type 1) within the object itself, determining fundamental transformation directions; secondary contradictions (Type 2) between system elements, such as when new tools collide with existing capabilities; and tertiary contradictions (Type 3) between old structures and new objects. An expansive transformation occurs when the object and motive are reconceptualized to embrace radically wider possibilities, representing a collective journey through the zone of proximal development (Engeström, 2001).

Table 1 illustrates how teachers' AI agent creation activity system manifests across these six elements. Using activity theory, this study diagnoses structural contradictions that explain why teachers' growing technical capabilities in Cycle 1 were not matched by sustained motivation, informing targeted interventions in Cycle 2.

**Table 1**
*Activity System Components in Teachers' AI Agent Creation*

| Components | Description in This Study |
| --- | --- |
| Subjects | In-service K-12 teachers participating in professional development workshops |
| Object → Outcome | Object: Addressing instructional challenges and student learning needs through AI-enhanced pedagogical approaches (TPACK)<br>Outcome: Teachers achieving professional identity transformation as AI pedagogical designers, capable of continuously creating context-specific intelligent teaching assistants aligned with curriculum goals and student learning needs |
| Tools | AI agent builder platform, prompt engineering techniques, knowledge base construction, design templates, AI Ethics |
| Rules | Workshop participation requirements, curriculum integration mandates, institutional assessment policies |
| Community | Peer teachers, workshop facilitators, technical experts, school administrators |
| Division of Labor | Roles shifting from teacher-as-learner (workshop training) to teacher-as-designer (agent creation) to teacher-as-implementer (classroom application) |

### Self-Determination Theory (SDT)

Self-Determination Theory distinguishes autonomous from controlled motivation, proposing that individuals possess three basic psychological needs: autonomy (feeling self-governed with ownership), competence (feeling capable and proficient), and relatedness (feeling connected with others) (Ryan & Deci, 2020). Environments satisfying these needs foster autonomous motivation and sustained engagement, whereas unmet needs lead to isolation, helplessness, and disengagement. Research demonstrates that when schools adequately support teachers' needs, teachers show greater commitment, willingness to embrace challenges, and persistence in complex practices such as technology integration (Chiu, 2022).

This study posits SDT as the guiding design framework for Cycle 2 intervention, with each need directly mapped onto specific redesign choices: autonomy frustration informed choice architecture allowing individual or collaborative modes and open topic selection; competence frustration informed progressive task scaffolding, embedded AI assistants, and discipline-specific templates; and relatedness frustration informed persistent community sharing and peer consultation channels. This need-to-design mapping operationalizes SDT, ensuring that structural redesign addresses the motivational deficits diagnosed through CHAT analysis.

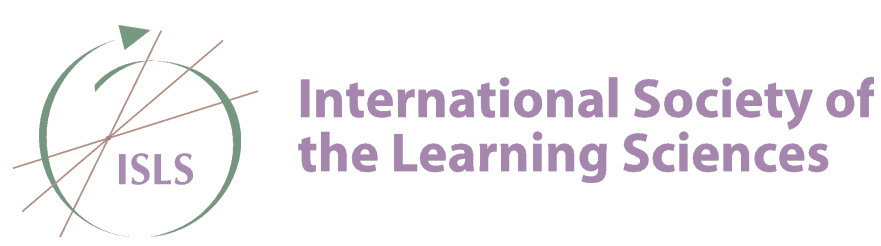

## Technological Pedagogical Content Knowledge (TPACK)

The TPACK framework conceptualizes teacher knowledge as dynamic interaction among content, pedagogy, and technology, extending Shulman's pedagogical content knowledge to encompass technological dimensions (Koehler & Mishra, 2009). Effective technology integration requires understanding how appropriate tools can teach particular content through effective pedagogical strategies (Koehler & Mishra, 2009). Celik's (2023) AI-enhanced TPACK incorporates intelligent technological knowledge dimensions alongside ethical considerations, recognizing that AI integration demands both technical proficiency and understanding of how these technologies interact with subject matter and pedagogical approaches. Complementing AI-TPACK, IPACK (Integrated Pedagogical and Content Knowledge for AI) specifically captures teachers' capacity to design AI-mediated instructional tasks and pedagogical agents (Tsai & Chai, 2012), providing a more proximal measure of agent creation capability.

AI-TPACK and IPACK operationalize "capacity" in this study: AI-TPACK indexes breadth of knowledge for AI integration, while IPACK captures the specific design competencies required for PA creation. This study employs both measures to track teachers' capacity development, treating TPACK enhancement as a necessary but insufficient condition—capacity provides the foundation while need satisfaction ensures sustained creative engagement.

# Method

## Research Design

This study employed a two-cycle formative intervention approach integrating Cultural-Historical Activity Theory and Self-Determination Theory. Drawing on formative intervention principles (Engeström, 2011), diagnostic findings from authentic practice contexts directly inform subsequent intervention redesign. However, unlike pure formative interventions where participants collectively identify contradictions, our research operated within district-mandated PD constraints, positioning researchers as interventionists who diagnose contradictions through systematic analysis (Figure 1).

Cycle 1 served as a diagnostic exploration, implementing baseline professional development followed by naturalistic observation without additional intervention. Eight-week platform behavioral tracking revealed that 87% of teachers ceased AI agent creation within three weeks. Semi-structured interviews with 28 purposively sampled teachers (selected through hierarchical clustering of behavioral profiles) enabled CHAT-based diagnosis of contradictions and SDT-based explanation of how systemic contradictions operated through psychological need frustration. Cycle 2 tested the intervention hypothesis through SDT-driven workshop. Pre-post surveys and interviews assessed whether need-supportive redesign could simultaneously enhance capacity and willingness for sustained AI agent creation.

**Figure 1**
*Two-Cycle Research Design and Theoretical Integration*

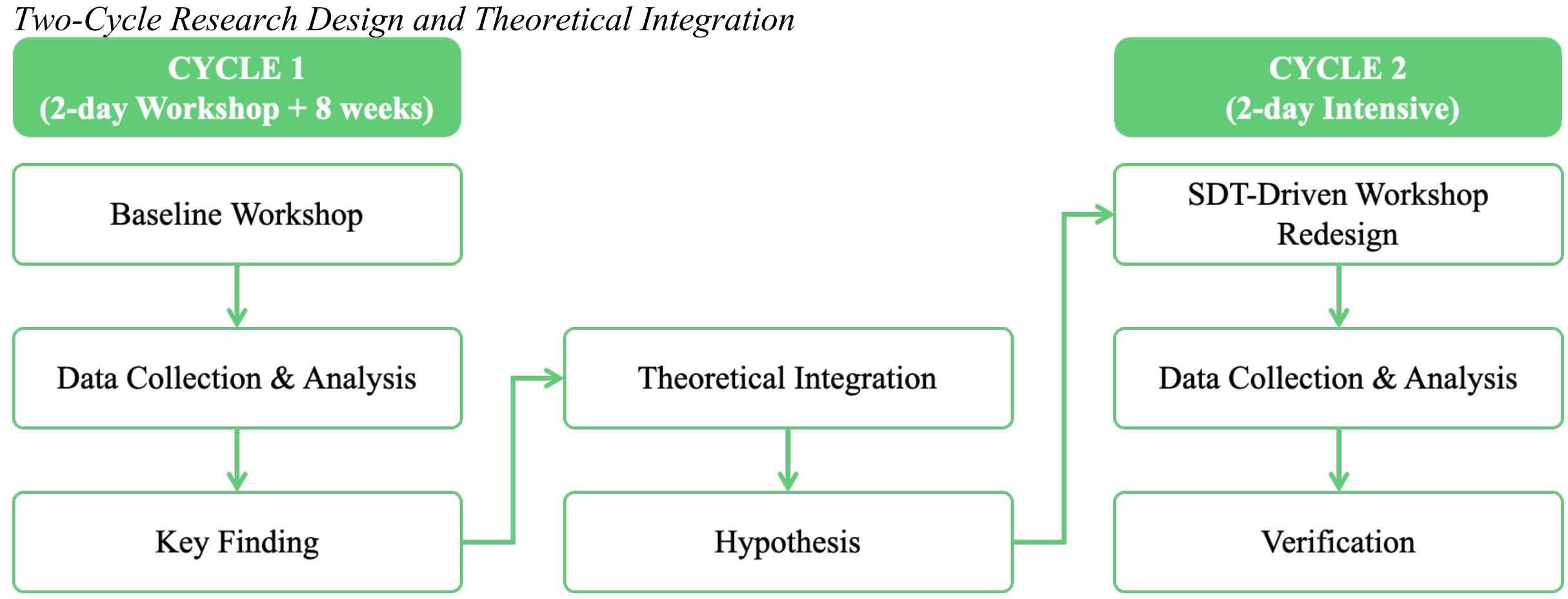

Figure 1 illustrates the research design. Findings from Cycle 1 revealed an unexpected phenomenon, prompting theoretical integration. These insights informed the redesign of the Cycle 2 intervention to test whether need-supportive environments could recouple the relationship.

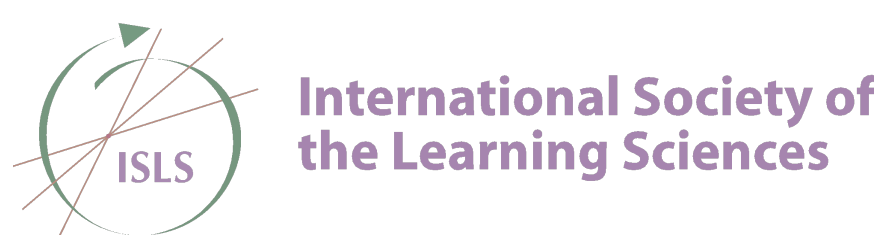

## Participant and Procedure

Participants were K-12 in-service teachers from a coastal city with well-developed educational infrastructure in Southern China—voluntarily enrolled in district-wide AI PD (teaching experience $M$=8.1-8.3 years, $SD$=5.1-5.3; gender distribution: male to female ratio = 1:2.9). The researchers functioned as interventionists who designed and facilitated both workshop cycles, drawing on systematic CHAT-based analysis to diagnose contradictions rather than positioning participants as co-investigators of their own practice.

Cycle 1 (N=218) implemented a two-day workshop in which teachers were introduced to generative AI concepts and AI agent design principles, then guided to design and build their own subject-specific educational applications using CocoFlow—an AI agent builder platform developed by the EdTech company COCOROBO Limited (Figure 2). CocoFlow's editing interface supports teachers in constructing pedagogical agents through prompt engineering, knowledge base creation via file upload, and workflow design via a visual drag-and-drop canvas. Teachers were expected to independently create at least one deployable AI agent aligned with their instructional subject and learning objectives, applying ICAP-informed pedagogical design principles (Chi & Wylie, 2014) to guide agent task design. Following the workshop, teachers entered an eight-week period without additional intervention but with full access to the platform, allowing naturalistic observation of their autonomous creation behaviors through continuous behavioral tracking. At Weeks 8–9, semi-structured interviews with 28 purposively sampled teachers (selected based on behavioral engagement profiles identified through hierarchical clustering) explored encountered barriers, motivational shifts, and workshop design perceptions.

Cycle 2 (N=26) was drawn from a separate cohort of teachers in the same district who voluntarily enrolled in a subsequent PD offering; the reduction from Cycle 1 reflects this separate-cohort design rather than attrition. Pre-tests were administered at the start of the two-day workshop (Day 1 morning), and post-tests were administered immediately upon workshop completion (Day 2 afternoon), enabling assessment of short-term capacity and willingness changes within the intensive intervention period. Cycle 2 retained core PD activities while integrating SDT-driven modifications targeting Cycle 1's diagnosed contradictions: autonomy support via choice architecture (individual/collaborative modes, topic selection); competence support via progressive scaffolding, embedded AI Q&A assistant, and discipline-specific templates; relatedness support via community sharing features and peer consultation mechanisms. Semi-structured interviews with 15 teachers explored AI-TPACK integration experiences and perceived need satisfaction.

## Instrument

Table 2 summarizes measurement instruments with administration timing and psychometric properties. Cycle 1 employed platform behavioral logs tracking seven variables over eight weeks, alongside semi-structured interviews with 28 purposively sampled teachers exploring contradictions. Cycle 2 added quantitative measures selected to operationalize the study's two core constructs: AI-TPACK (Celik, 2023) and IPACK (Tsai & Chai, 2012) surveys assessed capacity, and the Attitude toward AI Survey (Wang et al., 2024) assessed willingness to sustain AI agent use—all administered pre-post to capture change. The Basic Needs Satisfaction Survey (Li et al., 2025) was administered post-only to assess whether the redesigned workshop successfully created a need-supportive environment. All surveys used 7-point Likert scales with strong reliability (α=0.88-0.94). Interview transcripts were independently coded by two researchers (Cohen's κ=0.78).

**Figure 2**

*CocoFlow AI Agent Builder: Editing Interface*

The editing interface supports pedagogical agent construction through prompt engineering, knowledge base creation, and visual workflow design via drag-and-drop canvas operations.

**Table 2**
*Measurement Instruments Used Across Cycles*

| Cycle | Instrument | Construct Measured | Scale | Time | Reliability (α) | Source |
|---|---|---|---|---|---|---|
| Cycle 1 | Platform Behavioral Logs | Creation behaviors | Automated | 8 weeks | / | AI Agent builder |
| Cycle 1 | Semi-Structured Interviews | Barriers encountered during creation | 45-60 min audio-recorded | Week 8-9 (n=28) | / | Custom protocol |
| Cycle 2 | AI-TPACK Survey | AI-enhanced TPACK | 7-point Likert | Pre & Post | $\alpha$ = 0.89-0.94 | Celik (2023) |
| Cycle 2 | IPACK Survey | Capacity to create pedagogical agents | 7-point Likert | Pre & Post | $\alpha$ = 0.91 | Tsai & Chai (2012) |
| Cycle 2 | Attitude toward AI Survey | Willingness to use AI teaching assistants | 7-point Likert | Pre & Post | $\alpha$ = 0.88 | Wang et al. (2024) |
| Cycle 2 | Basic Needs Satisfaction Survey | SDT needs support | 7-point Likert | Post only | $\alpha$ = 0.90-0.93 | Li et al. (2025) |
| Cycle 2 | Semi-Structured Interviews | AI-TPACK integration; SDT needs satisfaction | 45–60 min audio-recorded | Post (n=15) | / | Custom protocol |

*Note.* All survey instruments underwent translation-back-translation procedures to ensure linguistic equivalence in the local contexts. Within the AI-TPACK survey, CK was assessed using discipline-neutral items addressing general content knowledge confidence, as the heterogeneous multi-subject sample precluded subject-specific measurement; dimensions unrelated to a teacher's specific subject (e.g., PK, Intelligent TK, Intelligent TPK) reflect cross-disciplinary AI integration knowledge applicable across all participants. The full AI-TPACK battery (including PK, CK, TK, TPK, TCK) is retained to document the breadth of capacity change; Table 4 highlights Intelligent TPACK and IPACK as the most proximal indicators of pedagogical agent creation capability.

## Data Analysis

**Cycle 1 Analysis.** Platform behavioral data underwent hierarchical clustering using Ward's linkage method with squared Euclidean distance on seven standardized variables (creation count, workflow count, edit count, browsing count, active days, session duration, feature utilization). Silhouette coefficients and dendrogram inspection determined optimal cluster solutions, with weekly activity frequencies aggregated to reveal temporal engagement trajectories. This behavioral profiling guided purposeful interview sampling to ensure representation across diverse engagement profiles.

Interview transcripts underwent iterative thematic analysis following Braun and Clarke's (2006) six-phase approach. Initial open coding identified barriers and motivational factors related to sustained creation. Axial coding organized codes into preliminary themes (technical, institutional, social-relational barriers), with selective coding consolidating findings into activity system contradictions. Two researchers independently coded 30% of transcripts, achieving substantial inter-rater agreement (Cohen's $\kappa$=0.78). CHAT contradiction analysis mapped identified barriers onto activity system elements (tools, rules, community, division of labor), with SDT framework applied to diagnose psychological need frustrations underlying motivational decline.

**Cycle 2 Analysis.** Paired-sample $t$-tests compared pre- and post-intervention scores on AI-TPACK, IPACK, and AI Attitude measures to assess intervention effectiveness. Correlation analyses examined relationships between post-intervention SDT need satisfaction scores and both capacity and willingness outcomes. Critically, partial correlation analyses controlled for baseline TPACK levels to test whether SDT need satisfaction uniquely predicted outcomes beyond pre-existing capacity, establishing need satisfaction as the associative language activating willingness. Effect sizes (Cohen's $d$) were calculated for all comparisons to assess practical significance. Interview data from 15 participants underwent thematic analysis focusing on two domains: TPACK integration experiences and SDT need satisfaction perceptions. Coding distinguished between high-SDT (above median) and low-SDT (below median) groups to identify boundary conditions for intervention effectiveness.

# Result

## Low Sustained Engagement Despite Workshop Completion (RQ1)

### Behavioral Engagement Patterns

Platform behavioral tracking over eight weeks revealed alarming engagement decline following the baseline workshop. Hierarchical clustering using Ward's linkage method identified three distinct profiles based on seven standardized variables. Silhouette coefficient analysis (0.62) confirmed optimal three-cluster solution: High-Creation-High-Browsing (10.6%, n=23) demonstrated sustained active creation with frequent browsing throughout the period; Low-Creation-High-Browsing (33.5%, n=73) focused primarily on browsing others' work with minimal creative output; and Low-Creation-Low-Browsing (56.0%, n=122) exhibited minimal platform engagement across all indicators.

Temporal analysis revealed a critical pattern: 87% of teachers ceased meaningful creation by Week 3. The Low-Creation-High-Browsing group showed concentrated activity in Week 1 followed by gradual decline, suggesting initial exploration without sustained practice. Even the High-Creation-High-Browsing group experienced a decline in creation in later weeks, though browsing remained active.

### Barriers Underlying Low Engagement

Semi-structured interviews with 28 purposively sampled teachers (distributed across three behavioral profiles) explored mechanisms underlying low engagement. Thematic analysis identified three barrier categories: Technical barriers (21/28) included platform functional limitations and insufficient subject-specific templates; Institutional barriers (18/28) revealed tensions between mandatory collaboration requirements and individual preferences, plus time pressures framing creation as additional burden; Social-relational barriers (16/28) centered on post-workshop isolation and uncertainty about seeking help. Notably, teachers attributed disengagement to environmental constraints rather than capability deficits, using phrases like "I could do it if..." followed by references to time, support, or tools, suggesting systemic contradictions rather than inadequate training.

## From Contradictions to Need Frustration (RQ2)

### CHAT Contradiction Diagnosis

Activity theory analysis identified three Type 2 secondary contradictions explaining teachers' motivational decline despite growing technical competence. Tool-Subject contradictions emerged where platform complexity exceeded real-world capabilities despite workshop competence, manifesting as cognitive overload when simultaneously managing technology, pedagogy, and content. Rule-Object contradictions surfaced where mandatory interdisciplinary collaboration conflicted with authentic creation needs; forced group tasks mismatched members' subject expertise and teaching contexts. Subject-Community contradictions centered on absence of sustained support—stark transitions from workshop environments rich with guidance to isolated individual practice.

Table 3 presents the complete contradiction framework with embodied manifestations. These contradictions were not implementation problems but structural tensions rooted in professional development models privileging researcher expertise over practitioner agency.

### Bridging Mechanism: Psychological Need Frustration

While CHAT diagnosed where problems existed, integrating Self-Determination Theory explained how contradictions inhibited motivation. We hypothesized that structural contradictions operated by thwarting basic psychological needs: Tool-Subject contradictions frustrated competence (repeated failures despite initial self-efficacy); Rule-Object contradictions frustrated autonomy (external control replacing volitional engagement); Subject-Community contradictions frustrated relatedness (professional isolation undermining belonging).

Interview analysis revealed this pattern: competence frustration manifested in "I can't get it right" clusters; autonomy frustration in external regulation language ("forced to," "required to"); relatedness frustration in professional loneliness descriptions. These need frustrations mediated the path from systemic barriers to motivational decline, suggesting Cycle 2 should directly target need satisfaction through activity system redesign.

## Intervention Verification—Recoupling Capacity and Willingness (RQ3)

### SDT-Driven Redesign and Synchronized Enhancement

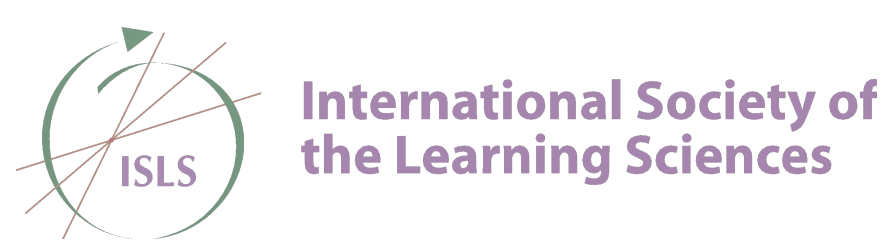

Cycle 2 implemented modifications targeting diagnosed contradictions through autonomy support (choice of topics/modes), competence support (progressive scaffolding, AI assistant, templates), and relatedness support (community sharing, peer consultation, continuous channels).

Paired-sample t-tests demonstrated significant intervention effects across multiple dimensions (Table 4). Most notably, AI-TPACK capacity ($d$ = 0.70), IPACK agent creation capacity ($d$ = 0.86), and AI attitude toward sustained use ($d$ = 0.58) all showed medium to large effect sizes. This synchronized enhancement suggests that SDT-driven redesign shaped both constructs concurrently, rather than improving capacity alone and expecting willingness to follow.

### Need Satisfaction as Explanatory Pathway

Cycle 2 successfully created need-supportive environments: overall SDT satisfaction ($M$ = 60.08, $SD$ = 11.74) substantially exceeded the theoretical midpoint of 53, with all sub-dimensions above their respective midpoints (Autonomy $M$ = 23.65, Competence $M$ = 18.75, Relatedness $M$ = 24.13). SDT showed strong positive associations with both capacity (AI-TPACK, $r$ = .865, $p$ < .001) and willingness (AI Willingness, $r$ = .732, $p$ < .001).

Partial correlation analyses controlling for all baseline measures revealed SDT need satisfaction as a robust predictor of post-intervention outcomes (Table 5). SDT showed strong positive associations with Intelligent TPACK ($r_partial$ = .715, $p$ < .001), IPACK agent creation capacity ($r_partial$ = .812, $p$ < .001), and critically, AI attitude toward sustained use ($r_partial$ = .521, $p$ = .032). All three SDT dimensions contributed significantly to overall need satisfaction. Notably, when controlling for baseline capacity, SDT's association with willingness (AI attitude) remained significant ($r_partial$ = .521, $p$ = .032), whereas post-intervention TPACK showed only a weak, non-significant association ($r_partial$ = .378, $p$ = .134). This pattern is consistent with our hypothesis that need satisfaction, rather than capacity enhancement per se, may function as a key associative pathway through which sustained willingness is activated.

**Table 3**
*Activity Theory Contradiction Analysis Framework*

| Contradiction Type | Embodied Manifestation |
|---|---|
| Type 1 Primary Contradiction (Within Object) | Value tension between AI's efficiency promise and educational goals emphasizing student cognitive development and critical thinking. |
| Type 2 Secondary: Subject - Tool | Platform's technical complexity and non-linear logic exceed teachers' real-world application capabilities despite workshop competence, generating cognitive overload. |
| Type 2 Secondary: Tool - Object | AI limitations (hallucinations, insufficient STEM knowledge) conflict with pedagogical accuracy demands, increasing teachers' review burden rather than reducing workload. |
| Type 2 Secondary: Subject - Community | Post-workshop isolation eliminates access to systematic tutorials, subject-specific templates, and real-time technical guidance available during training. |
| Type 2 Secondary: Rules - Division of Labor | Mandatory interdisciplinary group collaboration requirements mismatch actual subject expertise configurations, constraining authentic task completion. |
| Type 3 Tertiary: New vs. Old System | AI-empowered teaching paradigm emphasizing automation conflicts with traditional teaching culture valuing independent student thinking and effortful learning processes. |

**Table 4**
*Cycle 2 Pre-Post Intervention Effects on Capacity and Willingness*

| Dimension | *M_diff* | *SD* | *t* | *df* | *p* | Cohen's *d* |
|---|---|---|---|---|---|---|
| Pedagogical Knowledge (PK) | -4.42 | 6.59 | -3.43 | 25 | .002** | 0.67 |
| Content Knowledge (CK) | -0.89 | 2.46 | -1.84 | 25 | 0.078 | 0.36 |
| Intelligent TK | -2.42 | 4.03 | -3.07 | 25 | 0.005** | 0.60 |
| Intelligent TPK | -4.23 | 6.36 | -3.39 | 25 | 0.002** | 0.66 |
| Intelligent TCK | -2.04 | 4.44 | -2.34 | 25 | 0.028* | 0.46 |
| Intelligent TPACK | -4.58 | 6.50 | -3.59 | 25 | .001** | 0.70 |
| IPACK (Agent Creation) | -5.08 | 5.91 | -4.38 | 25 | <.001*** | 0.86 |
| AI Attitude | -2.00 | 3.45 | -2.95 | 25 | 0.007** | 0.58 |

*Note.* M_diff = Mean difference (post-pre); negative values indicate increase. *$p$ < .05, **$p$ < .01, ***$p$ < .001.

**Table 5**
*Partial Correlations Between Key Variables (Controlling for Baseline TPACK)*

| Variable | 1 | 2 | 3 | **4** | 5 | 6 |
|---|---|---|---|---|---|---|
| 1. SDT Need Satisfaction | - | | | | | |
| 2. Intelligent TPACK (post) | 0.715*** | - | | | | |
| 3. IPACK (post) | 0.812*** | 0.906*** | - | | | |
| 4. Perceived Autonomy Support | 0.917*** | 0.678** | 0.847*** | - | | |
| 5. Perceived Competence Support | 0.942*** | 0.594* | 0.818*** | 0.845*** | - | |
| 6. Perceived Relatedness Support | 0.899*** | 0.554* | 0.588** | 0.681*** | 0.784*** | - |
| 7. AI Attitude (post) | 0.521 | 0.378 | 0.533* | 0.567* | 0.521* | 0.358 |

*Note.* All correlations control for baseline measures: PK_before, CK_before, IntelligentTK_before, IntelligentTPK_before, IntelligentTCK_before, IntelligentTPACK_before, IPACK_before, AIAttitude_before, and Ethics_before. $^{*}p < .05$. $^{**}p < .01$. $^{***}p < .001$.

Qualitative Evidence: Transformed Experiences

Interviews with 15 Cycle 2 teachers illustrated need satisfaction transformation across three SDT dimensions. Competence: A low-SDT teacher described how progressive scaffolding converted failure into growth: "That success moment gave me confidence." Autonomy: A high-SDT teacher contrasted prior training with Cycle 2's choice architecture: "This autonomy greatly sparked my creative passion." Relatedness: Another high-SDT teacher described how community mechanisms replaced post-workshop isolation: "It feels like we're a team."

## Discussion

Cycle 1 revealed an unexpected phenomenon: despite completing a comprehensive two-day workshop, 87% of teachers ceased creating AI agents within three weeks. This was not a capacity problem but a systemic one. Activity theory diagnosis revealed that baseline TPD ignored systemic contradictions thwarting teachers' basic psychological needs: Tool-Subject contradictions generated competence frustration through platform complexity; Rule-Object contradictions produced autonomy frustration via mandatory collaboration; Subject-Community contradictions caused relatedness frustration through post-workshop isolation. Disengagement thus reflects rational responses to need-thwarting environments, challenging process-product logic (Opfer & Pedder, 2011) that assumes training activities automatically translate into sustained practice.

Cycle 2 demonstrated that SDT-driven activity system redesign simultaneously shaped both capacity and willingness—rather than improving one and expecting the other to follow. Three design principles emerged. First, diagnose systemic contradictions rather than surface barriers: mapping CHAT contradictions onto SDT needs provides precise intervention targets, preventing misallocated effort (Engeström, 2001; Ryan & Deci, 2020). Second, satisfy needs through structural redesign, not rhetoric: Cycle 2's effectiveness required concrete changes embedded in teachers' authentic practice contexts (Engeström, 2011), with boundary conditions indicating interventions succeed when tools match disciplinary needs and autonomy is substantive. Third, integrate capacity building with need satisfaction simultaneously: sequential approaches produce decoupling (Opfer & Pedder, 2011), whereas synergistic design—where scaffolds enhance both capability and satisfaction—enables sustained engagement.

## Implications and Future Direction

Theoretically, this study contributes CHAT-SDT functional integration as a replicable diagnostic logic extending beyond AI contexts, the low engagement paradox as a reframing of implementation failure from individual deficit to systemic contradiction, and associative evidence that need satisfaction is more strongly linked to sustained willingness than capacity gains alone. For administrators, protecting creation time (e.g., dedicated weekly periods for agent iteration) and granting tool selection agency prevent the autonomy deficits that drove 87% disengagement in Cycle 1. For designers, choice-based task portfolios, discipline-specific scaffolded templates, and low-barrier peer sharing spaces operationalize need-supportive PD concretely. For researchers, the CHAT-SDT framework offers diagnostic tools to trace why specific adoptions succeed or fail—moving from generic barrier attribution toward context-specific contradiction mapping and adaptive, system-aware intervention design.

Future research should examine long-term sustainability of teachers' AI agent creation behaviors and implementation effects in authentic classroom contexts, investigating how sustained design practices influence student learning outcomes and teachers' evolving pedagogical agency over extended periods.